\documentclass[twocolumn,superscriptaddress,aps,prl]{revtex4}
\usepackage{color}
\usepackage{amsmath}
\usepackage{amssymb}
\usepackage{graphicx}
\usepackage{subfigure}

\newcommand{\ds}{\displaystyle}

\begin{document}

\title{Spontaneously Broken Neutral Symmetry in an Ecological System}
\author{C. Borile}
\affiliation{Dipartimento di Fisica `G. Galilei' \& CNISM, INFN,
Universit\`a di Padova, Via Marzolo 8, 35131 Padova, Italy}
\author{M. A. Mu\~noz}
\affiliation{Instituto Carlos I de F\'isica Te\'orica y Computacional, Universidad de Granada, 18071 Granada, Spain}
\author{S. Azaele}
\affiliation{Institute of Integrative and Comparative Biology, University of Leeds,
Miall Building, Leeds LS2 9JT, United Kingdom}
\author{Jayanth R. Banavar}
\affiliation{Department of Physics, University of Maryland, College Park, MD 20742, USA}
\author{A. Maritan}
\affiliation{Dipartimento di Fisica `G. Galilei' \& CNISM, INFN,
Universit\`a di Padova, Via Marzolo 8, 35131 Padova, Italy}
\begin{abstract}
  Spontaneous symmetry breaking plays a fundamental role in many areas of condensed matter and particle physics. A fundamental problem in ecology is the elucidation of the mechanisms responsible for biodiversity and stability.  Neutral theory, which makes the simplifying assumption that all individuals (such as trees in a tropical forest) --regardless of the species they belong to-- have the same prospect of reproduction, death, etc., yields gross patterns that are in accord with empirical data.  We explore the possibility of birth and death rates that depend on the population density of species while treating the dynamics in a species-symmetric manner. We demonstrate that the dynamical evolution can lead to a stationary state characterized simultaneously by both biodiversity and spontaneously broken neutral symmetry.

\end{abstract}

\maketitle
Neutral models have been proposed to capture the statistical structure of tropical forests \cite{Hubbell01}. Even though the approach is highly debated \cite{Alonso06}, the neutral hypothesis has led to a general and fundamental framework to study both the statics \cite{Volkov05} and the dynamics \cite{Azaele06} of ecosystems using general tools borrowed from stochastic processes and non-equilibrium statistical mechanics. The fundamental assumption of neutral theory \cite{Hubbell01} is that within a trophic level any individual/organism behaves independently of the species it belongs to. In other words, the dynamics of the system is {unaffected} by interchanging/permuting species labels of individuals.  By using this extremely simplifying hypothesis many empirically measured statistical patterns can be well reproduced \cite{Chave,Volkov05,Azaele06}.  {Going one step further, a model can be \emph{symmetric} -but, strictly speaking, non-neutral-, a generalization of neutrality where the dynamics may depend, for instance, on the local or global density of individuals in a community, but no change occurs on the behavior of a population and on its effects on the others in the community upon switching two arbitrary species' labels \cite{Volkov05}}.  In this letter, we address the following issues: \textit{i)} Within a generalized neutral framework --allowing for intraspecific density-dependent demographic rates \cite{Chesson00}-- are species able to coexist in a stable way up to the temporal scale of speciation which eventually averts monodominance and extinction? \textit{ii)} Can this {generalized} neutral symmetry be spontaneously broken so that non-neutral behavior of species can emerge from an underlying {symmetric} dynamics?

In order to illustrate this, we consider a simple stochastic model, a variant of the (multi-species) voter model \cite{Liggett85,Durrett94}, defined as follows: at every vertex of a regular lattice of linear size $L$ in $d$ dimensions reside a \emph{fixed} number $M$ of individuals belonging to one of $S$ species. At every time step, an individual is picked at random and killed, and its place is filled by copying one of its neighbors selected according to a probabilistic rule to be defined in detail below. For illustration, let us consider a generic system of $S=4$ species and global dispersal where the neutral symmetry is not broken (see Fig. 1a). The fraction of each species' population fluctuates around the same average, $1/4$, and is statistically indistinguishable from the others. Also, at stationarity the four probabilities, $P_i(n)$, to find the $i$-th species with population $n$ are identical within statistical errors. In this case, the dynamics of the ecosystem is not changed by any permutation of species' labels; however, if each species has its own specific parameters for birth, death, dispersal etc., the dynamics is no longer symmetric. This explicitly broken symmetry makes the previous system of $S=4$ species behave in a completely different way (Fig. 1b). For instance, if a given species and the remaining are identified by distinct sets of parameters, the population fraction of one species fluctuates around a given average, $2/5$ in this case, whereas the other ones fluctuate around a different average, $1/5$. Even the probabilities $P_i$'s have distinct behaviors: three of them are identical and the fourth is different as shown in the left inset of Fig. 1b. Notice that the probability to find a species with $n$ individuals, $P(n)$, irrespective of the species identity, has a two-peak structure in the non-symmetric case.  Unlike the symmetric case, a non-symmetric model is necessarily characterized by a much larger set of parameters which make the approach unsuitable for understanding emergent phenomena (such as biodiversity). However, we will show in the present study that it is possible to define a symmetric theory from which non-neutral species' behaviors emerge naturally on appropriate temporal scales. This enables us to describe species-rich ecosystems with a parsimonious set of parameters which allows species to coexist without the overall symmetry characterizing the model. The idea that dynamical symmetry among species can be broken is not new in population biology. For instance, speciation can be interpreted as a form of bifurcation \cite{Stewart03}. However, here we introduce a new concept in community ecology which is borrowed from the statistical mechanics of phase transitions \cite{ZinnJustin02}, i.e. {\it spontaneously broken neutral symmetry}. As shown in Fig. 1c, when the symmetry of the model is broken spontaneously, species behave as in the non-symmetric case on time intervals shorter than a characteristic temporal scale, which will be calculated later on. On larger time scales, instead, species' identities can be swapped and eventually neutral dynamics is recovered. These large temporal scales are also comparable to those at which speciation can occur thereby sustaining biodiversity.  \begin{figure}[htbp]
 \subfigure{\resizebox{0.46\textwidth}{0.2\textheight}{\includegraphics{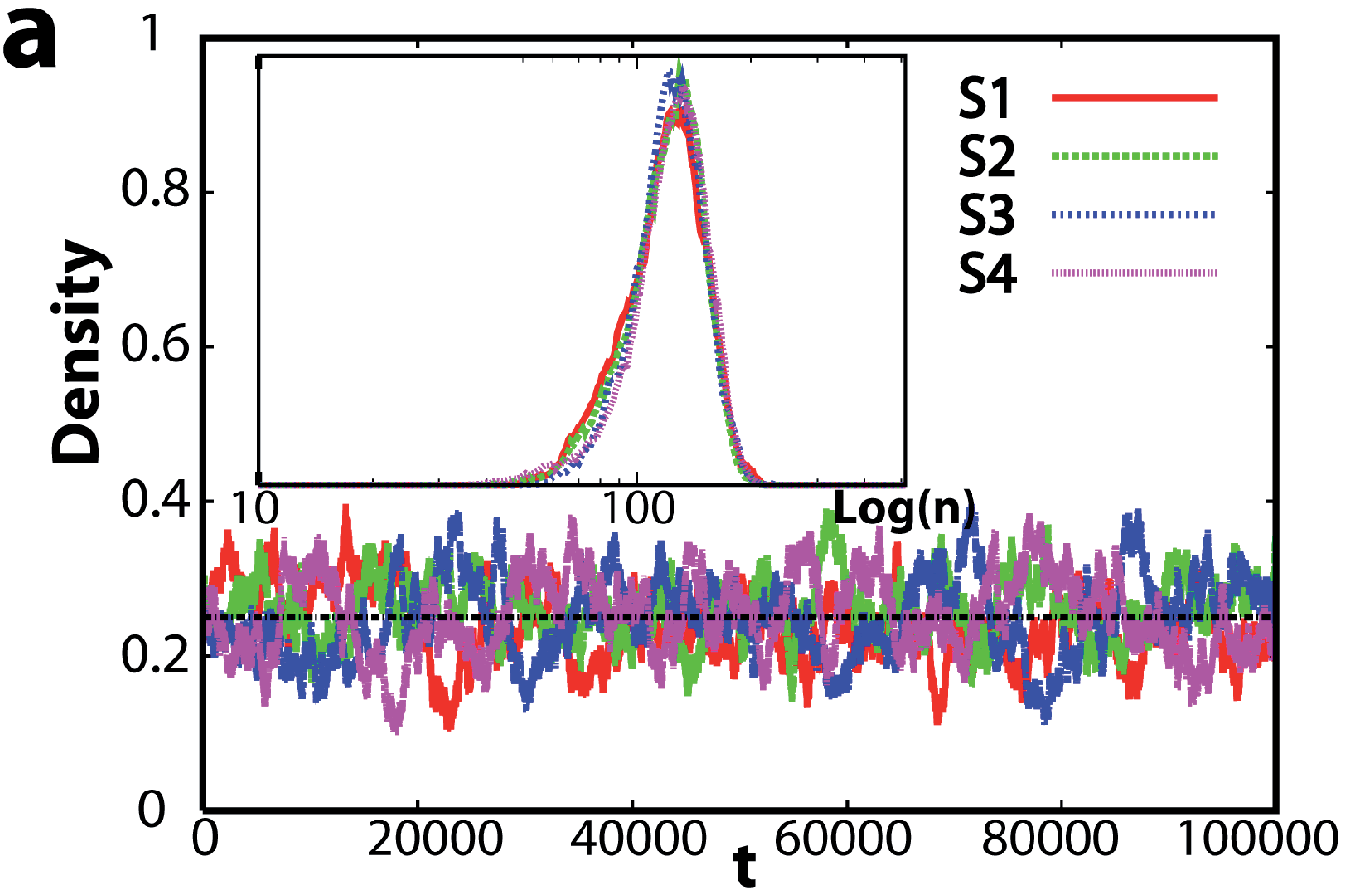}}}\\
 \subfigure{\resizebox{0.48\textwidth}{0.2\textheight}{\includegraphics{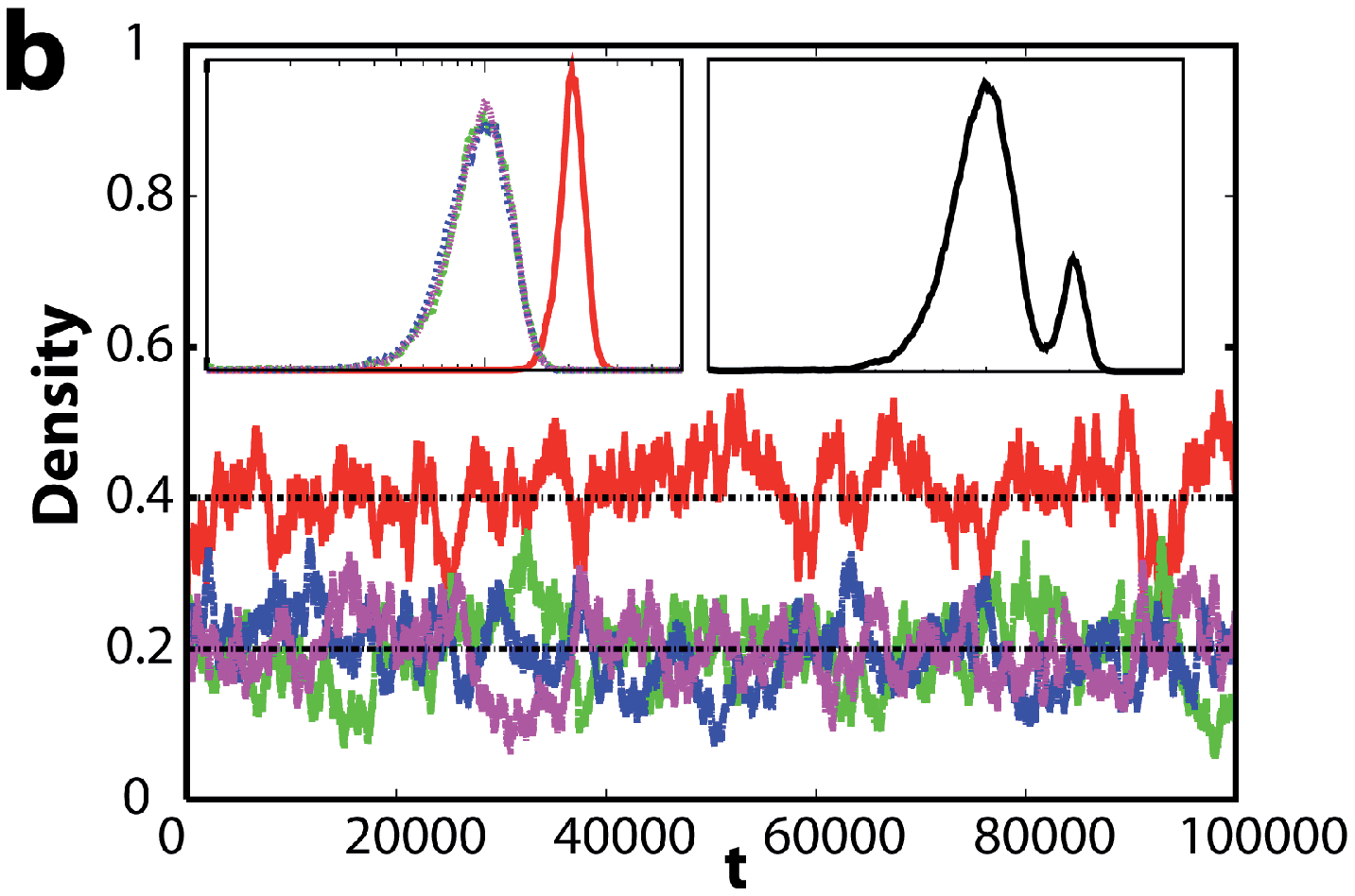}}}\\
 \subfigure{\resizebox{0.47\textwidth}{0.2\textheight}{\includegraphics{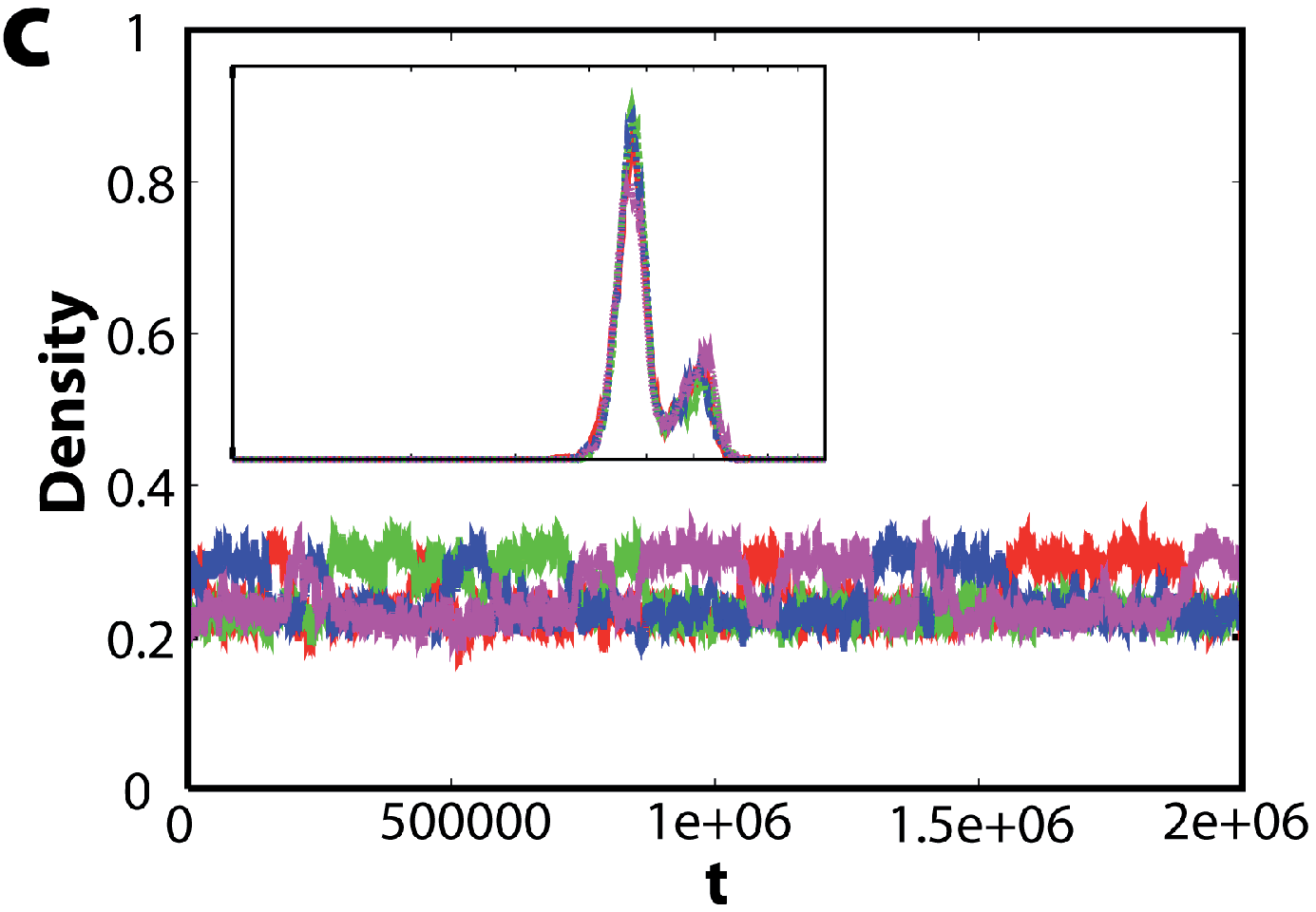}}}
\caption{Example of the evolution of a neutral ecological model with 4 species with global dispersal (see main text) for: (\textbf{a}) neutral symmetry. All the species are indistinguishable and fluctuate around the average value 1/4. In the inset (colors are the same as in the main picture) we show the probabilities $P_i(n)$, and the superposition is perfect within statistical errors, and (\textbf{b}) non-symmetric dynamics: species $1$ has a different set of birth and death rates with respect to the other three species, and fluctuates around an average density of $2/5$, while the others fluctuate around $1/5$. The probability $P_1(n)$ differs from the others, as shown in the left inset; in the inset on the right, the global probability $P(n)$ is shown. \textbf{c}) spontaneously broken neutral symmetry. Here the system behaves differently depending on the observation window of its evolution: for small time scales, the system appear non-symmetric, whereas, for longer time scales, the symmetry is recovered. Unlike case (b), all the species show a bimodal distribution.The probability $P(n)$ in this case superpose virtually exactly on the probabilities $P_i(n)$. The total population is $N=512$ individuals for the case a and b, and $N=2048$ individuals for c.}
\end{figure}

We turn now to the mathematical details of our model. Let $n_x^\alpha\geq 0$ the population at site $x$ of the $\alpha$-th species, where $\alpha = 1,\dots,S$, $S$ being the total number of species. Thus $\sum_{\alpha=1}^S n_x^\alpha = M$ holds for all $x$ and the total number of individuals in the whole community is $N=ML^2$. In the following, we shall also use the alternative variable $\rho_x^{\alpha}=n_x^{\alpha}/M$, the fraction/density of individuals of the $\alpha$-th species at site $x$. Suppose, that at time $t$, an individual belonging to a certain species $\gamma$ and at site $x$ is picked at random for removal. Then, call $\beta$ the species' label of the individual from one of neighboring sites of $x$, say $y$, selected to replace it. Note that the dynamics keeps the total population per site constant at every time step. Thus, a generic $n_z^\alpha$ evolves according to
\begin{equation}
n_z^\alpha \to n_z'^{\alpha} = n_z^\alpha + \delta_{x,z}\left(\delta^{\beta,\alpha}-\delta^{\gamma,\alpha}\right).
\end{equation}
The effective transition rate for this process is proportional to the population of the $\gamma$-th species at site $x$, $n_x^\gamma$, and to the population of the $\beta$-th species in the chosen neighboring site, $n_y^\beta$. Mathematically, this means that the probability of colonization is $P(n_x^\gamma\to n_x^{'\gamma})=K_{xy}^{\gamma\beta}n_x^\gamma n_y^\beta$. If the proportionality constant, $K_{xy}^{\gamma\beta}$, is chosen independently of the populations of species at $x$ and $y$ and independently of the kind of species involved, we get a voter like-model \cite{Liggett85, Durrett94} with neutral dynamics (the standard voter model has $M=1$, i.e. only one individual is allowed to live on each site). In this case, regardless of the initial conditions, an infinite size system would inexorably evolve towards a mono-dominant state, i.e. an absorbing state where only one of the $S$ species survives.  This is a trivial example of spontaneously broken neutral symmetry. In a more realistic perspective, however, different competing effects influence species interactions favoring or hampering colonization \cite{Molofsky99}, such as, for instance, the Janzen-Connell effect in tropical forests \cite{JC}, stating that the reproduction rate of a given species decreases with its local population size, or the Allee effect, a positive density dependence in a small density range \cite{Kot,Taylor05}. Altogether, these effects may result in an effective, in general non-linear and non-monotonic \cite{Molofsky99,Kot,Taylor05}, dependence on the population sizes, that we encode in the proportionality constant $K_{xy}^{\gamma\beta}$, now dependent, in principle, on the population sizes at both position $x$ and $y$. However, if the dynamics has to be neutral/symmetric then $K_{xy}^{\gamma\beta}$: \textit{i)} cannot depend explicitly on the species' labels $\gamma$ and $\beta$; \textit{ii)} can at best depend only on the densities of species $\beta$ and $\gamma$. Indeed, because the population of every site is fixed, we obtain the constraint $\sum_{\alpha\neq \beta,\gamma}\rho_x^{\alpha}= 1-(\rho_x^{\beta}+\ \rho_x^{\gamma})$ which is valid for every $x$ and plays an important role in the calculations. In order to keep the discussion simple, we consider the case $K_{xy}^{\gamma\beta}= K_{xy}(\rho_y^\beta)$, where $\rho_y^{\beta}$ represents the density of species $\beta$ at $y$ replacing one individual of species $\gamma$ at $x$.

In order to get some insight into the evolution of the ecosystem described above, following the standard approach for statistical mechanics systems, let us assume infinite dispersal or, equivalently, a well mixed system. This assumption - referred to as the mean field limit in the physics literature - is useful to simplify the treatment, while yet capturing the qualitative behavior of the model in any finite dimension. In this case, the description is simple since $\rho_x^\gamma=\rho^\gamma$ for all $\gamma=1,\dots,S$, the average birth rate of a generic species $\nu$ is proportional to $\rho^\nu(t)K(\rho^\nu(t))$ and the time derivative of $\sum_{\mu=1}^S \rho^\mu(t)$ has to vanish. Thus the evolution equation for the average density $\rho^\nu(t)$ can be derived by a standard Kramers-Moyal expansion \cite{Gardiner85} of the master equation of our system up to second order:
\begin{equation}
\begin{split}
\dot{\rho}^\nu =\ds N &\rho^\nu \Big [ (1-\rho^\nu) K(\rho^\nu)- \sum_{\mu\neq\nu} \rho^\mu K(\rho^\mu) \Big] \\
&+\Big \{\rho^\nu [(1-\rho^\nu) K(\rho^\nu)+ \sum_{\mu\neq\nu} \rho^\mu K(\rho^\mu) ]\Big\}^{\frac{1}{2}}\xi ,
\end{split}
\label{eq:Lang}
\end{equation}
where $\xi=\xi(t)$ is a Gaussian white noise $\delta$-correlated in time.
Focusing on the deterministic evolution, we set from here on $\xi(t)\equiv 0$. This is equivalent to neglecting fluctuations - of $\mathcal{O}(1/N)$ smaller than the deterministic term - in the analytical treatment. The simulations are performed by means of Gillespie's algorithm \cite{Gillespie77} considering directly the full master equation of the system.

The neutrality/symmetry of the dynamics is reflected in the stationary states obtained when $\frac{d}{dt}\rho^\nu(t)=0$. Note that the drift term on the rhs of Eq.(\ref{eq:Lang}) cannot be derived from a potential function and therefore the stationary states cannot be thought of as minima of an analytical function. However, regardless of the form of $K$, there are always $S+1$ steady states: one neutral-symmetric case, $\rho^\nu=1/S,\ \nu=1,2,\dots,\ S$, and $S$ mono-dominant situations where only one of the $\rho$'s is $1$ and the remaining ones are $0$. By using local stability analysis, one can prove that the mono-dominant states are stable only when $K(1)>K(0)$, whereas the condition $K'(\frac{1}{S})<0$ guarantees the stability of the symmetric coexistence. If the function $K(z)$ is linear, Eq. (\ref{eq:Lang}) has no other stationary stable solutions. However, in a more general non-linear case, new stable solutions can show up. It is this non-linearity that allows a spontaneous breaking of the neutral symmetry. The simplest situation of coexistence within a broken-symmetry scenario is obtained when a given species has density $\varphi > 1/S$ and all the other species have the same density $\zeta=(1-\varphi)/(S-1)<1/S$, which can occur in $S$ different ways. These densities correspond to stationary solutions of Eq.(\ref{eq:Lang}) if $K(\varphi)=K(\zeta)$ and are also stable when $K'(\zeta) < 0$ and $K'(\varphi)<-K'(\zeta)/(S-1)$.
We now discuss three paradigmatic cases.

\textbf{A}) $K=$ constant. This corresponds to the classic voter model \cite{Liggett85,Durrett94} (see fig. 2a). The deterministic evolution, given by eq.(\ref{eq:Lang}), is trivial because any initial value of the population of each species remains invariant across evolution. However, the stochastic dynamics leads to a mono-dominant state with only one surviving species, a trivial case of spontaneously broken neutral symmetry. For a finite system size, the time $\tau(N)$ to reach one of the $S$ absorbing states, starting from a random initial condition, scales as $\tau(N)\sim N^{\zeta}$ where $\zeta = 2$, as shown in fig. 3 (purple line) where $\log\tau(N)$ versus $\log N$ is plotted.

\textbf{B}) $K(z)=a(b-z)$ with $a,\ b>0$. This is a more interesting case (see fig. 2b) in which the colonization ability of a given species at some position decreases as its population --at the same position-- increases (negative density-dependence) and becomes zero when it reaches the maximum value $b$.  Therefore, abundant species are relatively not as effective in colonizing different regions compared to those with small populations.  The symmetric state is the stable stationary state of the deterministic evolution whereas the $S$ mono-dominant states are unstable. When the full stochastic dynamics is considered, the symmetric stationary state is reached, typically, after an initial transient (which depends on initial conditions). Once the stationary state is reached, it lasts for a typical time $\tau(N)\sim \exp\{\kappa N\}$, as shown in fig.3 (green line) and then the system evolves towards one of the $S$ mono-dominant states through a gradual extinction of species (observe that this exponential behavior is at variance with what happens in the $K=$ constant case where $\tau(N)\sim N^\zeta$). The constant $\kappa>0$ depends on the specific choice of $K(z)$.  The exponential behavior can be easily understood focusing on the limiting case of $S=2$ species, where a description in terms of a potential exists: Introducing a density-dependence in the Voter Model dynamical rule generates an effective potential in the equations of motion for $\rho^\nu, \nu=1,2$, that in the case of linear $K(z)$ discussed above has a minimum for $\rho=1/2$ (\cite{AlHammal05,Vazquez08,Castellano09,Dall'Asta10}). Thus, applying the well-known Arrhenius law and noting that the stochastic term is of order $1/N$ smaller than the deterministic part (see Eq. \ref{eq:Lang}), we recover the exponential behavior for $\tau(N)$. For time scales much smaller than $\tau(N)$ or for all times in the infinite size limit, $N\rightarrow\infty$, an active stationary state exists where all species symmetrically coexist. Therefore, negative density-dependence strongly enhances species coexistence.

We have calculated the relative species abundance (RSA) in the steady state, i.e. the probability, $P(n)$, to find a species with population $n$. The population $n^\nu(t)$ of the $\nu$-th species is followed for a sufficiently long time and the frequency, $P^\nu(n)\Delta n$, in each interval $(n,n+\Delta n)$ is recorded and the RSA is obtained as $P(n)=\sum_{\nu=1}^SP^\nu(n)/S$. In the neutral/symmetric case, the $P^\nu$ is independent of $\nu$ and the corresponding RSA is equivalent to those in figure 1a. Note that at variance with the $K=$ constant case --where the RSA is not well defined as a consequence of the lack of metastable active states \cite{Dickman02}-- in the case $K(z)=a(b-z)$, (see Fig. 1a), we obtain a mode, as typically found in the RSA of several tropical forests \cite{Hubbell01,Volkov05,Azaele06} and other ecosystems \cite{Volkov07}.

\textbf{C}) $K(z)$ has the '\textit{S}' shape shown in fig. 2c  \cite{Molofsky99,Taylor05} in order to satisfy the stability conditions for a broken symmetry scenario given above (this particular shape is for convenience, but it is also valid for $K(z)$ of the generic cubic form $K(z)=az^3+bz^2+cz+d$ with suitably chosen coefficients; note that a cubic non-linearity in the density-dependence is usually called a {\it Nagumo}
 term and is employed to describe populations experiencing the Allee effect \cite{Kot}).
Here the broken-symmetry coexistence is the stable stationary state of the deterministic evolution. Turning on the stochastic dynamics --after an initial transient-- the system reaches one of the $S$ stationary states of the deterministic dynamics with broken symmetry. Again, on a typical time scale $\tau(N)\sim \exp\{\kappa' N\}$ there is gradual extinction of species till, one gets a mono-dominant situation. Once more, the constant $\kappa'>0$ depends on the specific choice of $K(z)$.
When the system is in a broken-symmetry case, the species whose density fluctuates around the average $\varphi>1/S$ interchanges with one of the $S-1$ species fluctuating around the average $\zeta=(1-\varphi)/(S-1)<1/S$ on time scales $\tau_{switch}(N) \sim \exp\{k_{s} N\}$. Thus in a finite system, $N<\infty$ and on a time scale $\gg \tau_{switch}(N)$ the ecosystem looks neutral/symmetric, i.e. species behave like they were interchangeable. However, for time scales $\ll \tau_{switch}(N)$ or for all times within an infinite system, $N=\infty$, the neutral symmetry is spontaneously broken and the ecosystem looks as if species were not all interchangeable. We have calculated the probability, $P^\nu(n)$, that the $\nu$-th species has population $n$ on a time scale smaller than $\tau_{switch}(N)$ so as to exhibit the characteristics of a broken-symmetry state. The results are indistinguishable from  those of the case where there is no neutral symmetry (Fig 1b), in which we run the model with two different functions $K(z)$ depending on species label: for $\nu=1$ we set $K(z)=K_1(z)=a_1-b_1z$ with $a_1=3$ and $b_1=2$, while for $\nu=2,3,4$ we set $K(z)=K_2(z)=a_2-b_2z$ with $a_2=2.5$ and $b_2=1.5$. The RSA for the spontaneous symmetry breaking case calculated for time scales $\gg \tau_{switch}(N)$ is displayed in the inset of fig. 1c where two peaks appear, showing that one of the species behaves differently from the others. In a more general pattern of spontaneous symmetry breaking, one can have up to $S$ distinct $P^\nu$'s producing a $S$-peak RSA. Multiple peaks would be resolved in the RSA depending on the width and separation of the peaks: this scenario is consistent with some recent studies on several different ecological communities \cite{VanNes12} pointing out the possibility of a multimodal distribution of $P(n)$ in real systems.

In conclusion, we have shown that a simple non-equilibrium microscopic model for a general $S$-species ecological community driven by a density-dependent but otherwise completely neutral/symmetric dynamics --\textit{i.e.} the dynamic rules governing the stochastic microscopic process are insensitive to the species' labels-- can show a rich and stable heterogeneous biodiversity even at very long times. The striking fact is that species can behave distinctly by spontaneously breaking the neutral symmetry.

\begin{figure}[htbp]
 \centering
 \resizebox{0.39\textwidth}{0.16\textheight}{\includegraphics{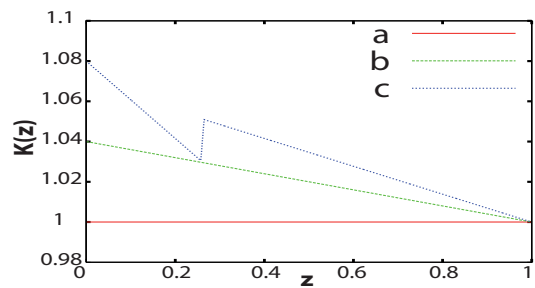}}
\caption{(Color online) \textbf{a)} (Red solid line) $K(z)\equiv 1$, corresponding to the standard Voter Model with many species. \textbf{b)} (Green dashed line) $K(z)=a(b-z)$: This definition of the function $K(z)$ makes the symmetric state stable against perturbations, and the monodominant states unstable, provided $a>0$. \textbf{c)} (Blue dotted line) $K(z)$ allowing $S$ stable stationary states where the neutral symmetry is spontaneously broken by one of the $S$ species.}
\label{fig:K}
\end{figure}

\begin{figure}[htbp]
 \centering
 \resizebox{0.43\textwidth}{0.16\textheight}{\includegraphics{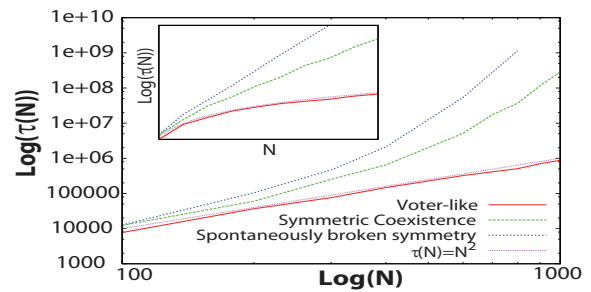}}
\caption{(Color online) Mean time to extinction $\tau(N)$ for the three different definitions of $K(z)$ in Fig. (\ref{fig:K}), calculated in the mean field approximation and plotted in Log-Log scale varying $N$ from $N=100$ to $N=1000$. For $K=const.$ (red solid line), $\tau(N)\sim N^\alpha$ with $\alpha \simeq 2$ (red dotted line)  as expected for a Voter-like model, while the two cases of $K=b-az$ (green dashed line), where we chose $a=0.04,\ b=1.04$, and $K(z)$ allowing for a spontaneous breaking of the neutral symmetry (blue dotted line) show an exponential behavior $\tau(N)\sim e^{kN}$. In the inset, we show the same plot in a Log-Linear scale, to emphasize the exponential growth.}
\end{figure}

\end{document}